# Autonomous Adaptive Security Framework for 5G-Enabled IoT

Habtamu Abie and Sandeep Pirbhulal
Norwegian Computing Center, PO Box 114, Blindern, 0314 Oslo, Norway
Emails: abie@nr.no, sandeep@nr.no

**Abstract.** In IoT-based critical sectors, 5G can provide more rapid connection speeds, lower latency, faster downloads, and capability to connect more devices due to the introduction of new dynamics such as softwarization and virtualization. 5G-enabled IoT networks increase systems vulnerabilities to security threats due to these dynamics. Consequently, adaptive cybersecurity solutions need to be developed for 5G-enabled IoT applications to protect them against potential cyber-attacks. This task specifies new adaptive strategies of security intelligence with associated scenarios to meet the challenges of 5G-IoT characteristics. In this task we have also developed an autonomous adaptive security framework which can protect 5G-enabaled IoT dynamically and autonomously. The framework is based on a closed feedback loop of advanced analytics to monitor, analyse, and adapt to evolving threats to 5G-enanled IoT applications.

## 1 Introduction

5G Networks have various advantages such as fast speed, low latency, scalability, etc compared with previous 1G, 2G,3G and 4G technologies [Liu 2019]. 5G introduces new dynamics due to the use of NFV, network slicing, and SDN [Hellaoui 2020]. 5G-enbaled IoT systems are prone to security risks due to high dependability on softwarization and virtualization. Therefore, adaption of security services needs to be considered with respect to new 5G-IoT dynamics. This adaptation can be done both during use of security services and before since the 5G-based architecture allows such adaptation.

Looking in the lens of 5G-enabled IoT dynamics can offer a systematic way of understanding, characterizing, quantifying, and managing cybersecurity. Risk-managements and decision-making questions can automatically be answered by quantifying cybersecurity [Xu 2020]. In addition, a dynamic risk assessment can be built to make the solution risk-aware and quantify cybersecurity, and secure identity management and execution environments can be integrated using 5G-SIM card. In this work we follow the work by Hellaoui et al. [Hellaoui 2020] to adapt end-to-end security, security level related to cryptographic keys during security establishment phase, and data origin authentication.

This paper presents the development of an autonomous adaptive security framework for 5G-enabaled IoT which can protect 5G-enabaled IoT systems dynamically and autonomously. The framework specifies new adaptive strategies of security intelligence with associated scenarios to meet the challenges of IoT and 5G dynamics and characteristics. It is based on a closed feedback loop of advanced analytics to monitor, analyse, and adapt to evolving threats to 5G-enanled IoT applications.

## 2   Background and Related Work

Neto et al. [Neto 2022] survey the essential mission of security requirements of PPDR (Public Protection and Disaster Relief) systems, describe the state of the art of the most used technologies, and discuss their evolution to 4G and 5G cellular networks. 5G is useful for various IoT applications but it has severe security issues raised in several research publications [Ahmad 2018, Arfaoui 2019, Moudoud 2021]. This section will briefly describe the fundamental changes in 5G security, IoT security and 5G-enabled IoT security.

### 2.1   5G Security

Tele-communications play an essential role in citizens' well-being and on most of the Critical Infrastructures (CIs) operations. In CIs, along with an efficient connection for communication connectivity, the other most important factors are speed, data rate, availability, and latency. 5G provides more rapid connection speeds, lower latency, faster downloads, and capability to connect more devices.
Nevertheless, it is expected that the 5G can be used for illegal actions driven by several reasons, including espionage and cyberwarfare, state-sponsored political motives, and adversaries, among others [John 2022].
Next-generation communication technologies radically change how we communicate by presenting various new capabilities, connections, and services in multiple sectors. In 5G, two promising technologies are a) network function virtualization and b) end-to-end network slicing empowering 5G networks for efficient and dynamic deployment of network. Cybersecurity and Resilience of CIs are vital factors that can decrease vulnerability, reduce the effects of threats and their cascading consequences, and accelerate mitigation procedures [Maria 2021]. 5G is connecting all aspects of life by bringing a digital revolution that needs increased service availability and security utilizing various technologies [Ahmad 2018].

5G security risks are mainly due to 5G networks having various stakeholders based on multiple security specifications [Moudoud 2021]. The key challenges of 5G and Next Generation Mobile Networks (NGMN) are stated as follows [Ahmad 2018] and as depicted in Figure 1:

- *User plane integrity*: The user plane carries the end-user's data. Protecting the integrity of user plane is essential for smooth communications between 5G devices and networks.
- *Flash network traffic and signaling storms*: In 5G networks, several devices are connected, which may yield flash network traffic by compromising vulnerable devices.  In an environment where a huge number of devices are connected it may pose a considerable threat to the 5G signaling network.

- *Roaming security: Secure roaming in 5G networks is important for mobile operators to protect subscribers and network infrastructures.*
- *Decentralized security: 5G communication based on blockchain technology requires more traffic routing points thus it will be important to create trustworthy communication.*
- *DoS attacks on end-user devices and infrastructures: DoS attack could make devices or networks or infrastructures inaccessible to the envisioned users by flooding the target with traffic.*
- *Compatibility: 5G devices will have more bandwidth requirement thus it may have high pressure on the current security monitoring tools to process in a smooth manner.*
- *IoT device security: Many IoT devices are designed in a way that they do not have built-in security, therefore device security might be considered separately.*
- *Lack of early encryption in connection process: Receiving unciphered information of the 5G device by an attacker allows it to launch cyber-attacks.*

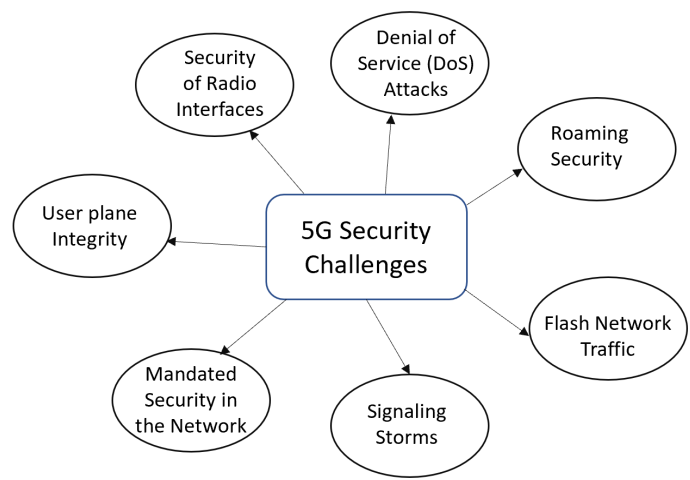

Figure 1 - Fundamental 5G Security Challenges

### 2.2 IoT Security

There has been exponential growth in deploying the IoT in various applications including healthcare, transportation, defense, industry, etc. It is no surprise that numerous recent security attacks worldwide are IoT-based in which the hackers manipulate vulnerable technology as an initial phase toward compromising any CI connected with an IoT network. These attacks are evident in various infrastructures such as energy, industry, vehicular, and healthcare since IoT technologies are an integral part of these infrastructures. Yet, in some sectors where IoT is deployed at the end-user side, e.g., smart

44

homes, security attacks can be miscalculated since all possible attack paths cannot be analyzed [Loannis 2018].

Such a broad application of IoT requires institutions to raise remarkable awareness of system security. Any susceptibility can be the reason for failures or attacks in the system; several people can be affected by these risks. For example, traffic lights might not work causing roadway casualties; burglars could turn off a home security system. There are some nodes in IoT related to health monitoring or human protection, and their security can be essential for individuals' lives. As per a report [Bayern 2018], $1.24 billion was IoT security market in 2017; however, it is expected to be around $9.88 billion by 2025 which shows the significance of IoT security.

Assuring IoT security is tricky because many intelligent nodes have limited computational power, and it is not possible to apply complex security functions because they require more energy and resources [Zhou 2018]. Therefore, IoT-based systems are believed to have more vulnerabilities than their counterpart [Katrenko 2022]. Several IoT systems have security vulnerabilities for the following reasons:

- Lack of computational power for effective built-in protection
- Inadequate access control
- Limited resources for appropriately experimenting and enhancing firmware protection
- Lack of periodic patches and updates due to restricted budgets and technological constraints of IoT devices
- Users cannot update their devices, restricting susceptibility patching
- Insufficient security from physical threats since attackers may reach near enough to the chip or attack the nodes using radio signals.

IoT is widely used in critical sectors having direct or indirect societal and social impacts [Krishna 2017]. On one hand, IoT can improve individual's daily lives, but various challenges may restrain its evolution, including interoperability, design problems, security, business challenges, etc., [Alkhalila 2017]. For example, DDoS attacks (October 2016) on company Dyn caused the inaccessibility of several websites containing Twitter, GitHub, etc. [Kamran 2021]. This attack is performed using the botnet that includes numerous IoT nodes, including child monitors, printers, IP cameras, etc. Although IoT devices have been an essential part of everyday lives, people may not prefer using this technology if proper security and privacy measures are not in place. The high computational cryptographic algorithms cannot be used with hardware and software programs, considering the power-draining issue. Zhou et al. analyzed and discussed IoT security challenges and indicated that intrusion detection could be a potential solution [Zhou 2018]. Tkaczyk et al. presented software design patterns to resolve interoperability issues in IoT ecosystems [Tkaczyk 2018]. The most critical IoT security challenges are weak passwords, poor IoT device management, secure signing, insecure update deployment, human aspects in IoT, compromised privacy and protection, insecure authentication and authorization, cryptojacking, lack of compliance and prevention and identification run parallel [Sectrio 2022].

### 2.3    5G-enabled IoT Security

5G-enabled IoT networks increase systems vulnerability to security threats [Hajar 2020]. 5G-enabled IoT has several applications, including remote surgery, industry 4.0, etc. In a 5G-enabled IoT communication environment, device and user communicate through the Internet which faces different cybersecurity issues, including DDoS, malware, man-in-the-middle, etc. Consequently, cybersecurity solutions need to be developed for 5G-enabled IoT applications to protect them against cyber-attacks [Mohammad 2020]. Several innovative solutions for securing 5G-enabled IoT communication are required, such as security key management, intrusion detection, root cause analysis, device /user authentication, access control, etc. The security and privacy challenges for 5G-enabled IoT communication include lack of robust security schemes, openness of the network and privacy of sensitive data.

## 3    Challenges and Security Mechanisms

A recent report commissioned by the European Union [NIS 2019] suggests that 5G networks are prone to various security risks due to dependability on softwarization and virtualization. This section describes cybersecurity strategies for cybersecurity scenarios to enhance autonomous adaptive security for 5G-enabled IoT.

### 3.1    Dynamics

Hellaoui et al. [Hellaoui 2020] argue that new dynamics are introduced by 5G due to softwarization, NFV, network slicing, SDN, and service customization. IoTs are dynamic by nature due to collaborative communications, resource constraint and usage. The dynamics nature or characterization of 5G-enabled IoT can be summarized as follows [Hellaoui 2020]:

- *IoT dynamics collaboration:* IoT can join and leave group communications due to resource constraint and usage (services provisioning) which makes IoT environments and collaborations dynamic. For instance, a node can join a communication group to provide or consume services and when not interested in the services it can leave the group. This kind of communication represents IoT dynamics.

- *5G dynamics:* Softwerization of the network in 5G introduces dynamic. This network softwerization and programability is useful for IoT. NFV, SDN and network slicing are also key technologies in 5G which introduce dynamic. This paper investigates these dynamics of 5G-enabled IoT to offer autonomous adaptive security.

### 3.2    Automation and Autonomy

Automation usually refers as the capability of a system to identify and mitigate anomalies without human efforts. Autonomy is the ability to learning, anticipate, adaptive and respond to dynamic situations like threats, and to evolve as the environment around it changes. For instance, self-adaptive systems autonomously identify, analyze and mitigate threats by adapting their structure and behavior at runtime to cope with uncertainties in their environment and their dynamic nature.

Parasuraman et al. [Parasuraman 2000] and Cazorla et al. [Cazorla 2013] propose application of automation based on these categories: data collection, data analytics, decision making, and measures implementation. The automation of these functions will be considered during implementation and to move from automation to autonomous, the observation, interpretation, decisioning, and action tasks performed by human need to be sufficiently automated.

Boutaba et al. [Boutaba 2021] and Parasuraman et al. [Parasuraman 2000] argue the significance of autonomous systems and the interaction with their operator based on different policies and strategies. The application of automation types and levels need to be considered during the implementation phase of this task to benchmark moving automation to autonomous. Prasad et al. [Prasad 2021] elaborate how IT operators can integrate with AI analytics, predication capabilities and augmentation of processes and knowledge. Redyuk et al. [Redyuk 2021] present different scenarios and datasets which can be used to validate automation. The framework will adapt this type of automation.

### 3.3 Adaptation

Abie et al. [Abie 2012] define adaptation with reference to adaptive security and how it is important for critical sectors (healthcare). Adaptive 5G-enabled IoT security solutions must thus learn, adapt, and adjust continually using closed feedback loop solutions to stay ahead of their adversaries. In our case the security solutions must adapt during the security establishment phase and running phase considering the IoT and 5G dynamics to mitigate security relevant uncertainties and to provide timely countermeasures to complex and frequent attacks.

### 3.4 AI/ML closed feedback loop: Closed loop dynamic

Gomes et al. [Gomes 2021] argue that intent-driven closed feedback loop using AI/ML can be used to create an end-to-end automation framework for autonomous network management. Schöning et al. [Schöning 2022] argue that closed-loop control systems can provide high precession for real-time applications using AI/ML analytics.

Prasad et al. [Prasad 2021] present the following key capabilities in network operations that closed-loop automation enables anomaly detection, smart alerts, prediction and root cause identification.

There exist different applications of CLA. Samal et al. [Samal 2021] describe a closed-loop approach for autonomous drones using neural networks. Caesar et al. [Caesar 2021] proposed a technique for autonomous driving focusing on goal-based planning, planning metrics (Common metrics and Scenario-based metrics) and closed-loop evaluation. Boutaba et al. [Boutaba 2021] describe how the integration of AI/ML and closed loop can used for service orchestration in 5G networks. Xie et al. [Xie 2020] describe AI-driven closed-loop service assurance architecture for 5G systems. Daryanavard et al. [Daryanavard 2020] presented AI/ML based solutions for multi-layer networks using feedback and control engineering concepts.

Our closed feedback loop AI is inspired by the above work but focuses more on providing continuous learning, improving detection and intelligence based on context-aware to recognize and incorporate new and changed knowledge and make decisions. This will allow building a deep and interconnected understanding to autonomously analyze and adapt to intrusion and contexts changes and enhancing cyber intrusion detection and intelligent monitoring by continually optimizing as the AI algorithms improve with data and experience.

### 3.5 Dynamic Risk Assessment

Cybersecurity industries, research and academic institutions, and government agencies developed several risk assessment workflows for evaluating risks [Behbehani 2022]. However, such approaches lack in managing dynamically changing circumstances and cannot adapt their countermeasures and preferences to change in transformations internally or externally [Naumov 2016]. Another significant problem in risk management includes automated information scarcity and uncertainties to allow satisfactory risk computation [Zhang 2017] [Zhang 2018]. Dynamic risk assessment approaches may assist in the realistic assessment of the operational risk in complex environments [Yu 2016] [Ahmadi 2020]. The statistic approaches tend to be incomplete and may not fully address the latest and more diverse risks emerging from the widespread application; thus, the assessment results may be inaccurate. They may also be unable to predict known or unknown attacks due to varying conditions [Huang 2017].

A dynamic risk assessment has great potential for 5G enabled IoT systems because it provides the capability to monitor direct and indirect communication, thus useful for operational risk estimation. A dynamic risk assessment framework which continuously monitors and analyses in any changing environment in real-time to identifying and reducing intentional and unintentional risks during systems development and operations. Bayesian Networks and fuzzy logics can be useful for developing an efficient dynamic risk analysis because of their potential to model probabilistic information [Berenjian 2016] [Zhang 2018], thus can be useful for 5G-enabled IoT applications. It can significantly improve 5G-enabled IoT security by addressing threats in any changing environment in different real-time scenarios.

### 3.6 5G SIM Card for Identity Management and Trusted Execution

5G-SIM card can be used to build a more secure identity management & access control layer for both telecom operators and end users. It can also be used to build secure isolated tamper proof execution environments for running sensitive applications, such as IoT device introspection, threat detection and payment. Depending on the calculated risk exposure of a given application/workload, the autonomous adaptive security module can in fact decide to deploy/make use of the stronger security guarantees provided by using the 5G-SIM card as a root of trust and/or a more secure execution environment. This strategy will build secure identity management and access control for 5G-enabled IoT systems and develop secure isolated tamper proof execution environments for running sensitive 5G-enabled IoT applications using 5G-SIM Card. It will provide a proof-of-concept for secure identity management and access control and tamper proof execution environments for running sensitive 5G-enabled IoT applications using 5G-SIM Card.

### 3.7 Adaptive Authentication

5G networks leverage a flexible and dynamic infrastructure using softwerization and virtualization for dynamic scenarios. Arfaoui et al. [Arfaoui 2019] explain the necessity of dynamic environments for 5G-enabled IoT applications because it offers overall overview analysis and adaptive mitigating measure, and proposed solutions to meet these requirements. Rezaeibagha et al. [Rezaeibagha 2022] developed cryptographic mechanism for mobile edge computing based IoT, which allows authentication between different nodes and edges.

To start a secure communication, an adaptive authentication procedure needs to adaptively check the identities of different devices in 5G-enbled IoT [Mohammad 2020, Montaño-Blacio 2021]. In this task, we consider a widely adopted scenario [Hamdi 2014, Hellaoui 2020, Mauro 2015] related to data origin authentication including digital signatures from devices' agnostic hardware secure element (e.g., eUICC/eSIM) to ensure trust in 5G-enabled IoT data producers.

### 3.8 End-to-end security

In our end-to-end 5G-enabled IoT scenario we consider the IoT device (software, firmware, hardware, applications, etc.), 5G infrastructure (radio network, the core network, etc.), MEC (Multi-access edge computing) and cloud (connects the IoT network to the outside world, platform for managing the IoT devices) that connects the IoT network to the outside world. During the implementation phase we will consider the adopted scenarios in [Hellaoui 2020, Mauro 2015] and extend them with adaptive security intelligence using AI/ML techniques.

### 3.9 Adaptive Anomaly Detection and Prevention

The adaptive anomaly detection and prevention model is critical for 5G-enabled IoT applications for monitoring, predicting, and verifying abnormal activities, and providing protection against the intrusions. Adversary may try to set up pernicious nodes to launch some hazardous attacks. IoT botnets may also attack the communication through installing malware in IoT devices. Although there exist several anomaly detection techniques, but they don't consider varying and dynamic nature of 5G-enabled IoT for improving anomaly detection and prediction. Therefore, in this task we consider the adaptive anomaly detection and prevention techniques for our developed framework which addresses the dynamic characteristics of 5G-enbled IoT to predict attacks in advance and verification of insufficient protection.

Vávra et al. [Vávra 2021] develop neural network based adaptive techniques to protect 5G-enabled IoT systems against DDoS and FDIA attacks. Motivated by their efforts we developed adaptive anomaly detection and prevention model which applies closed loop AI techniques (e.g., RNN, CNN, DNN) to introduce the detection and prevention intelligence. The AI techniques will play main role in detecting and predicting cyber threats and attacks in the 5G-enabled IoT networks through continuous learning and improving detection and intelligence to recognize and incorporate new and changed knowledge obtained from real-time feedback and make decisions and then adapt the dynamics accordingly.

## 4 Proposed Framework for Autonomous Adaptive Security

In this section, we propose an autonomous adaptive cybersecurity framework for securing IoT networks and applications based on distributed infrastructure as shown in Figure 5, a framework for security-related data collection, analytics and prediction of incidents and provision of response and mitigation measures autonomously. The framework is based on a closed feedback loop of advanced analytics to monitor, analyse, and adapt to evolving threats to 5G-enanled IoT applications.

The framework uses (i) adaptive cybersecurity services, (ii) three layers as in [Moudoud 2021], (iii) a closed feedback loop AI for prediction, detection mitigation model.

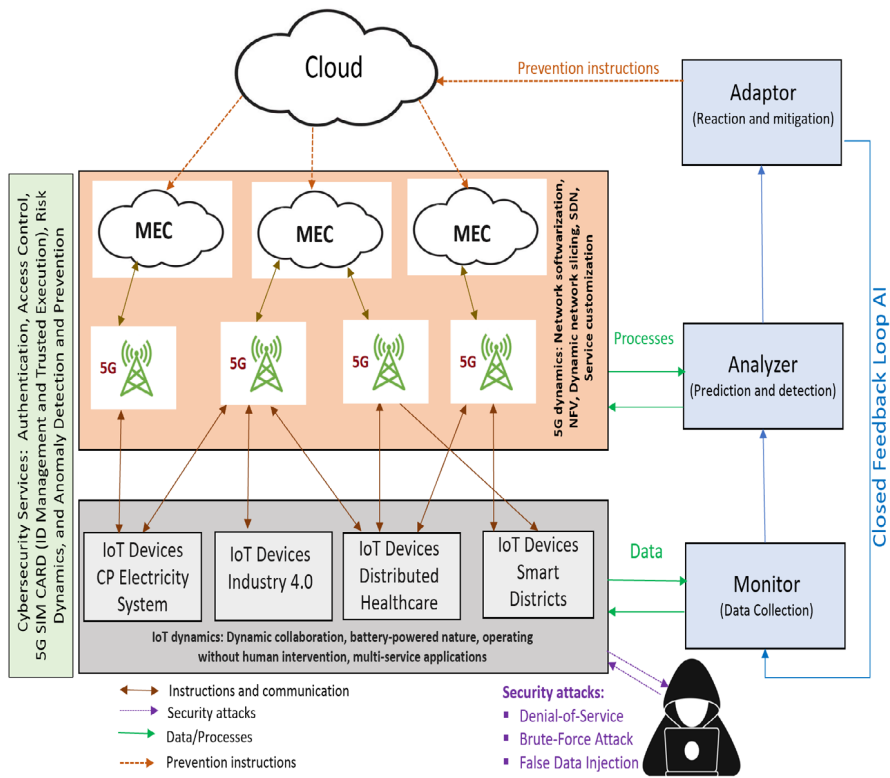

Figure 2 - Autonomous Adaptive Security Framework for 5G-enabledIoT

### 4.1 The Adaptive Cybersecurity Services

The adaptive Cybersecurity Services as described in Sections 3 & 4 include Authentication, Access Control (end-to-end security), 5G SIM CARD (ID Management and Trusted Execution), Risk Dynamics, and Anomaly Detection and Prevention.

### 4.2 The Devices Access Layer

The devices access layer collects and transfers data to the MEC layer via the 5G gateways in real time with fast speed and low latency. The data from different IoT devices is transferred via 5G gateways providing protection of sensitive and critical data.

### 4.3 The MEC layer

Due to the resource constraint nature of the IoT devices our developed framework uses the MEC layer to process and analyze the collected data from the device layer.

However, MEC suffers from security challenges despite processing a high amount of traffic and data leakage issues due to sharing data among multiple devices. For these reasons, we propose to implement adaptive prediction and detection model to mitigate FDIA and DDoS attacks, which will enable the MEC layer to orchestrate data transmission, communicate control information, and update protection measures.

### 4.4 The Cloud Layer

The cloud layer will manage big data collected by the MEC layer. To reduce security and privacy attacks it is required to isolate the shared clouds. In this regard one possible solution is use virtual private cloud which has low potential to be compromised. To limit access to resources and data, access control can play pivotal role for security of users.

### 4.5 Closed Feedback Loop AI Model

The Closed feedback loop AI model in our proposed framework will be used for prediction, detection, and mitigation of security attacks. It will improve detection and intelligence based on context and metadata to recognize and incorporate new and changed knowledge and make decisions. The continuous closed feedback loop AI will provide a deep, interconnected understanding of the 5G-IoT networks to autonomously analyze and adapt to intrusion and contexts changes. Thus, enhancing cyber intrusion detection and intelligent monitoring by continually optimizing as the AI algorithms improve with data and experience.

Initially, the Closed Feedback Loop AI analyzes IoT data to predict potential security threats, carefully plan and critically reason to understand the dynamic 5G-IoT environments and make decisions on mitigation measures in a closed feedback loop. The outer closed feedback loop between the 5G-enabled IoT-Edge-Cloud continuum and the Closed Feedback Loop AI brings an efficient way to monitor, control, analyze, predict, and anticipate cybersecurity threats to the 5G-IoT-Edge-Cloud continuum. The Closed Feedback Loop AI sends the mitigation measures to the 5G-IoT-Edge-Cloud continuum for updates and continues to monitor, analyze, and adapt to the dynamic security threats closing the outer feedback loop.

Lastly, the self-learning feature can be achieved through the internal close feedback loop of each AI techniques which is the subject later investigation.

The proposed Closed Feedback Loop AI comprises three models: the Monitor model which collects data, the Analyzer model which analyzes collected data for prediction and detection of security threats, and the Adaptor model which reacts and mitigates attacks. Each model uses AI technique such as RNN, CNN, or DNN. We call this "Closed Feedback Loop AI".

### 4.6 The Monitor Model

The Monitor Model continuously collects, aggregates, filters, and reports contextual information collected from the 5G-IoT environments. In 5G-enabled IoT system, a data cleansing is important to correct and remove inaccurate information to ensure data quality which can likely result in more accurate attack detection. Duplicate data need to be removed, and only representative samples can be kept which can reduce unnecessary calculations for the prediction and detection of attacks. Yet, dynamic context monitoring is essential for 5G-enabled IoT to constantly monitor the changing environments.

For the monitor model most popular monitoring frameworks such as "OpenStack, Open Source MANO, OSM MON, VNFM, Self-monitoring Scripts, OpenStack Ceilometer + Aodh [Xie 2020] can be considered. We agree with the authors [Boutaba 2021] that adaptive algorithms are useful for efficient monitoring of frequencies for specific slices and metrics.

### 4.7 The Analyzer Model

The Analyzer (Prediction and Detection) Model analyses the collected data for detecting and predicting cyber risks and planning mitigation measures. In 5G-enabled IoT there exist heterogeneous devices. Therefore, it is important to convert collected data from devices into an understandable and unified format. Activities related data can be classified into different categories as specified in [Boutaba 2021] to allow an efficient response to security threats alongside data protection.

To extract knowledge and inference for predicting and detecting security attacks, we propose RNN, CNN, or DNN. Security decisions can dynamically be adapted using advanced estimation and predictive models.

### 4.8 The Adaptor Model

The Adaptor Model (Reaction Model) adapts security measures for addressing constantly evolving threats in real time. Building a secure 5G-enabled IoT system requires an interconnection between all security mechanisms. For instance, devices authentication, end-to-end security (access control) and privacy measures must be considered together. The Adaptor Model must adapt security mechanisms (anomaly detection and prevention, security level, encryption algorithms, security protocols). This can be determined by the data activity categories mentioned earlier. For instance, a deleting activity can be automatically blocked, and device's access privileges be prohibited. For the other two remaining activities verification of the corresponding state is required.

## 5 Validation Use Cases

This study also highlights the potential use cases for this task in collaboration with Sykehuset Innlandet (5G-enabled IoT for IT-OT integration with the perspective of the healthcare domain) and SINTEF (smart grid protection considering cybersecurity in 5G-enabled applications).

### 5.1 Healthcare

Smart Healthcare: Our existing ASSET (adaptive security for the smart Internet of Things in eHealth) platform [Berhanu 2013] collects the data from the Raspberry Pi and Shimmer motes, and stores data in the cloud for remote consultation. In [Pirbhulal 2022], we developed the DT of the ASSET platform to enable the modelling and monitoring of human activities and biosignals using digital models and near real-time synchronisation with the physical system. This real-time DT of the smart health systems with 5G-enabled IoT capabilities can be used for improving the quality of care, and automated prediction of security risks.

### 5.2 Smart energy grids

Figure 3 shows a possible use case for autonomous secure 5G-IoT-Enabled advanced metering infrastructure. Advanced metering infrastructure (AMI) is a component of smart energy grids (smart meters and collectors). It can nicely be modelled as 5G-IoT-Edge-Cloud continuum as depicted in Figure 3, IoT devices representing smart meters, 5G gateways and MEC as collectors/aggregators, and the Cloud as the head-end system (HES) to manage smart meter data. To validate such a use case can contribute to improve grid reliability, security and efficiency in communications and advanced applications.

The main area in which the use case clearly demonstrates progress beyond the state of the art includes autonomous 5G-IoT-AMI adaptive security approach for device-edge-cloud continuum through continuous security-related data collection from IoT devices/smart meters, analytics, prediction and anticipation of incidents and provision of response and mitigation measures autonomously. The model knows exactly the right action to take, at the right time, to contain an in-progress attack. We are looking for use case partners to develop and work together on this use case.

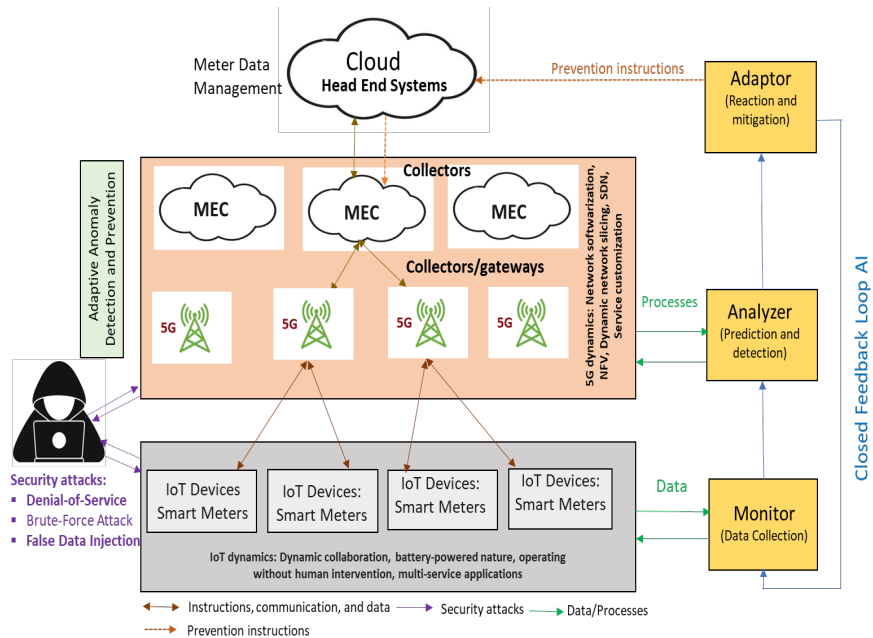

Figure 3 – Autonomous Framework for Secure 5G-IoT-Enabled Advanced Metering Infrastructure

### 5.3 Industry 4.0

Industry 4.0: Industry 4.0 revolutionizes physical and digital technologies to acquire essential insights about their operations [Andrew 2022]; these insights can play a vital role in improving performance. 5G provides connectivity of Industry 4.0 with improved network latency and enables a range of critical applications, e.g., autonomous driving. 5G-enabled IoT provides better speed and QoS that can be helpful in real-time orientated application and how that application has to communicate continuously to the ever-changing conditions.

### 5.4 Smart district

Smart Cities: 5G technology supports the advancement of digitalization in smart cities in different locations and infrastructures such as land, roads, rails and water. Our development in [Sodhro 2019], developed a resource-efficient technique for IoT-enabled green and sustainable smart cities. Our resource-efficient solution can be merged with developed 5G-enabled IoT framework to provide a balance of energy-efficiency and security in smart cities.

## 6     Conclusions and Future Work

This paper describes the developed autonomous adaptive security framework for 5G-enbaled IoT with the new adaptive strategies and specifications for cybersecurity scenarios. This paper argues that security services must be adapted to meet the 5G-IoT dynamic characteristics. Therefore, we developed an autonomous adaptive security framework for 5G-enabled IoT for security-related data collection, analytics and prediction of incidents and provision of mitigation measures autonomously. The framework is based on a closed feedback loop of advanced analytics to monitor, analyse, and adapt to evolving threats to 5G-enanled IoT applications. This paper also describes 4 use cases based on discussions with NORCICS partners which need more detailing and elaborations.

In our future work we will detail, implement, and simulate some of the main components and validate them in suitable use cases.

**Acknowledgment**

This work has received funding from the Research Council of Norway through the SFI Norwegian Centre for Cybersecurity in Critical Sectors (NORCICS), project no. 310105, and basic institute funding at Norwegian Computing Center (Norsk Regnesentral), RCN grant number 194067.

1626. [Boutaba 2021] Raouf Boutaba et al. (2021). AI-driven Closed-loop Automation in 5G and beyond Mobile Networks. In Proceedings of the 4th FlexNets Workshop on Flexible Networks Artificial Intelligence Supported Network Flexibility and Agility (FlexNets '21). Association for Computing Machinery, New York, NY, USA, 1–6. https://doi.org/10.1145/3472735.3474458
27. [Berhanu 2013] Berhanu, Y., Abie, H., and Hamdi, M. (2013). A testbed for adaptive security for iot in ehealth. In Proceedings of the International Workshop on Adaptive Security, pages 1–8
28. [Pirbhulal 2022] Pirbhulal, S., Abie, H., & Shukla, A. (2022, June). Towards a Novel Framework for Reinforcing Cybersecurity using Digital Twins in IoT-based Healthcare Applications. In 2022 IEEE 95th Vehicular Technology Conference:(VTC2022-Spring) (pp. 1-5). IEEE
29. [Andrew 2022] Andrew Ross, https://www.information-age.com/5g-is-the-heart-of-industry-4-0-13970/, July 2022
30. [Sodhro 2019] Sodhro, A. H., Pirbhulal, S., Luo, Z., & De Albuquerque, V. H. C. (2019). Towards an optimal resource management for IoT based Green and sustainable smart cities. Journal of Cleaner Production, 220, 1167-1179.
31. [Parasuraman 2000] Raja Parasuraman et al. (2000), A Model for Types and Levels of Human Interaction with Automation, 286 IEEE Transactions on Systems, Man, and Cybernetics—Part A: Systems and Humans, Vol. 30, No. 3, 286-97, May 2000
32. [Prasad 2021] Sharath Prasad et al. (2021]. An introduction to closed-loop automation: Benefits, challenges, use cases, and architectures for implementing closed loop automation systems, 29 September 2021, https://developer.ibm.com/articles/an-introduction-to-closed-loop-automation/
33. [Cazorla 2013] L. Cazorla, C. Alcaraz, and J. Lopez (2013). Towards Automatic Critical Infrastructure Protection through Machine Learning. In: Luiijf E., Hartel P. (eds.) Critical Information Infrastructures Security. CRITIS 2013. Lecture Notes in Computer Science, vol. 8328. Springer, Cham.
34. [Redyuk 2021] Sergey Redyuk et al. "Automating Data Quality Validation for Dynamic Data Ingestion." EDBT (2021). Series ISSN: 2367-2005, pp 61-72, 10.5441/002/edbt.2021.07
35. [Abie 2012] Habtamu Abie and Ilangko Balasingham. 2012. Risk-based adaptive security for smart IoT in eHealth. In Proceedings of the 7th International Conference on Body Area Networks (BodyNets '12). ICST (Institute for Computer Sciences, Social-Informatics and Telecommunications Engineering), Brussels, BEL, 269–275.
36. [Gomes 2021] Gomes, Pedro Henrique et al. (2021). "Intent-driven Closed Loops for Autonomous Networks." J. ICT Stand. 9 (2021): 257-290.
37. [Schöning 2022] Julius Schöning et al. (2022). "AI for Closed-Loop Control Systems---New Opportunities for Modeling, Designing, and Tuning Control Systems." arXiv preprint arXiv:2201.06961 (2022).
38. [Samal 2021] K. Samal, M. Wolf and S. Mukhopadhyay, "Closed-loop Approach to Perception in Autonomous System," 2021 Design, Automation & Test in Europe Conference & Exhibition (DATE), 2021, pp. 463-468, doi: 10.23919/DATE51398.2021.9474243.
39. [Caesar 2021] H. Caesar et al. (2021). nuplan: A closed-loop ml-based planning benchmark for autonomous vehicles. arXiv preprint arXiv:2106.11810.
40. [Daryanavard 2020] Sama Daryanavard and Bernd Porr. (2020). "Closed-loop deep learning: Generating forward models with backpropagation." Neural Computation 32.11 (2020): 2122-2144.